%% file: main.tex
\documentclass[sigconf,screen,nonacm]{acmart}

\usepackage{booktabs}
\usepackage{tabularx}
\usepackage{multirow}
\usepackage{subcaption}

\setcopyright{rightsretained}
\acmYear{2026}
\copyrightyear{2026}
\acmConference[Preprint]{Preprint}{2026}{}
\acmISBN{}
\acmDOI{}
\title{Quantifying Ethereum Energy Consumption via Network Mapping}
\subtitle{A Node-Level Estimate from Crawlable Attributes}

 \author{Yahn Costa Hackspacher}
 \orcid{0009-0007-7218-2033}
 \affiliation{%
   \institution{FIZ Karlsruhe -- Leibniz Institute for Information Infrastructure}
   \city{Karlsruhe}
   \country{Germany}
 }
 \email{Yahn.Costa-Hackspacher@fiz-Karlsruhe.de}

  \author{Cornelius Ihle}
 \orcid{0000-0002-3994-5218}
 \affiliation{%
   \institution{University of G\"ottingen}
   \city{G\"ottingen}
   \country{Germany}
 }
 \email{ihle@gipplab.org}

  \author{Vasundhara Shaw}
 \orcid{0000-0002-5824-7191}
 \affiliation{%
   \institution{FIZ Karlsruhe -- Leibniz Institute for Information Infrastructure}
   \city{Karlsruhe}
   \country{Germany}
 }
 \email{Vasundhar.Shaw@fiz-karlsruhe.de}

 \author{Dennis Trautwein}
 \orcid{0000-0002-8567-2353}
 \affiliation{%
   \institution{ProbeLab Analytics OÜ}
   \city{Talinn}
   \country{Estonia}
 }
 \email{dennis@probelab.io}

 \author{Geerd-Dietger Hoffmann}
 \orcid{0009-0005-4948-395X}
 \affiliation{%
   \department{Department of Computer Science}
   \institution{University of Potsdam}
   \city{Potsdam}
   \country{Germany}
 }
 \email{didi@green-coding.io}

 \author{Bela Gipp}
 \orcid{0000-0001-6522-3019}
 \affiliation{%
   \institution{University of G\"ottingen}
   \city{G\"ottingen}
   \country{Germany}
 }
 \email{gipp@uni-goettingen.de}

 \author{Moritz Schubotz}
 \orcid{0000-0001-7141-4997}
 \affiliation{%
   \institution{FIZ Karlsruhe -- Leibniz Institute for Information Infrastructure}
   \city{Karlsruhe}
   \country{Germany}
 }
 \email{moritz.schubotz@fiz-karlsruhe.de}



\begin{document}
\renewcommand{\topfraction}{0.92}
\renewcommand{\bottomfraction}{0.7}
\renewcommand{\textfraction}{0.07}
\renewcommand{\floatpagefraction}{0.82}
\renewcommand{\dbltopfraction}{0.92}
\renewcommand{\dblfloatpagefraction}{0.82}
\setcounter{topnumber}{4}
\setcounter{dbltopnumber}{2}
\setcounter{totalnumber}{8}
\setlength{\textfloatsep}{7pt plus 2pt minus 3pt}
\setlength{\floatsep}{6pt plus 2pt minus 2pt}
\setlength{\intextsep}{7pt plus 2pt minus 3pt}
\setlength{\dbltextfloatsep}{8pt plus 2pt minus 4pt}
\setlength{\dblfloatsep}{6pt plus 2pt minus 2pt}
\raggedbottom

\input{sections/01_abstract}
\maketitle
\input{sections/02_introduction}
\input{sections/03_related}
\input{sections/04_method}

\input{sections/05_results}
\input{sections/06_discussion}
\input{sections/07_conclusion}
\bibliographystyle{ACM-Reference-Format}
\bibliography{references}
\clearpage
\input{sections/08_appendix}

\end{document}

%% file: sections/01_abstract.tex
\begin{abstract}
Ethereum's electricity use fell by about 99.95\% after the network
moved from proof of work to proof of stake. Service providers still
need to report how much energy their operations use, e.g.\ under
the EU Markets in Crypto-Assets Regulation (MiCAR). Existing
estimates either apply one typical wattage to every node or start
from aggregated monitoring counts. Both approaches ignore information
that nodes already advertise on the peer-to-peer network: which
client software they run, whether they use ARM or x86 hardware,
where they are hosted, and whether they act as validators.

We combine crawls of the consensus and execution layers, assign a
wattage to each peer from those attributes using published
measurements, and estimate the remaining incomplete peers with a
Random Forest. On 6{,}934 peers from two Nebula crawls (19 and
22~June~2026), the nodes that the crawler can reach sum to 415~kW,
or 3.63~GWh if that draw were held for a year. Applying the same
Lighthouse$+$Nethermind x86 wattage to every one of those peers
instead yields 431~kW. Using the observed attributes therefore
lowers the total by 3.9\%, mainly because nodes at Hetzner and
other non-AWS clouds draw less than that home-desktop figure. AWS
then accounts for 15.6\% of watts from 12.2\% of peers, and
validator-flagged nodes account for 31.4\% of watts from 25.7\% of
peers.

Our snapshot of 415~kW is about 46\% of the Cambridge Centre for
Alternative Finance (CCAF) estimate of about 0.90~MW. Both
approaches use about 60~W per node, so the gap is mostly how many
nodes each estimate includes. The assignment rules cover 3{,}110
peers and the forest estimates the other 3{,}824. 
On held-out labeled peers with client, architecture, and OS hidden, the forest's mean absolute error against the rule wattage is 4.3~W.
Twenty-four-hour measurements on a gaming desktop differ from the
predictions. After subtracting a 33~W idle
graphics card that Ethereum clients do not need, both differences
fall to about 19\%.
\end{abstract}

%% file: sections/02_introduction.tex
\section{Introduction}
\label{sec:introduction}

Ethereum's Merge in September 2022 replaced proof of work (PoW),
long criticised for its energy use \citep{leible2019review},
with proof of stake (PoS) and cut network electricity use by
about 99.95\% \citep{ethereum_merge}. Pre-Merge annualised
estimates reached about 94~TWh
\citep{digiconomist_eth,ccaf2026report}. After the Merge,
security comes from staked ETH and the remaining energy cost is
the wall power of machines that run consensus-layer (CL) and
execution-layer (EL) clients. The Cambridge Centre for
Alternative Finance (CCAF) places current demand near 0.90~MW,
or 7.87~GWh/year \citep{ccaf2026report}.

Dialable nodes run different client pairs on ARM boards
\citep{cortes2023home} and on high-memory cloud instances. Regulation
(EU)~2023/1114 on Markets in Crypto-Assets (MiCAR) requires energy
information in white papers (Articles~6(1), 19(1), 51(1)) and
updated figures from service providers (Article~66(5))
\citep{micar2023}. The delegated sustainability standards make
annual energy, in kilowatt-hours, the mandatory indicator.
Greenhouse-gas and renewable-share fields apply only once that
energy exceeds 500{,}000~kWh/year, and claiming renewable
electricity or offsets does not reduce the energy figure itself
\citep{micar2025rts}.

Existing post-Merge models either apply one weighted wattage to
every participant \citep{ccri2022merge} or start from aggregated
headcounts \citep{cbnsi_methodology,pankovska2024electricity}.
Neither uses metadata that the peer-to-peer protocols already
expose. The libp2p \texttt{AgentVersion} string encodes the client,
processor architecture, and operating system
\citep{libp2p_identify_spec,geth_node_config,lighthouse_releases}.
The Ethereum Node Record (ENR) carries subnet bitvectors that
indicate whether a node is acting as a validator
\citep{EIP778,eth2_p2p_spec,sutton2022validators}. Without
attributing each peer's IP address to a hosting provider, the
virtualisation-aware cloud model of
\citet{pankovska2024electricity} also cannot be applied instance
by instance \citep{lei2021best,pankovska2024electricity}.

The reachable-set total used below is the sum of estimated wall
power over peers that Nebula can dial on mainnet, that appear on
both CL and EL crawls, and that receive a rule label or a forest
imputation. NAT-only DHT clients, one-layer rows, and churn
between crawl sessions sit outside that set
\citep{antswatch,kiffer2025thirtysix}. It is a snapshot lower
bound, not a fleet count.

We address the following two questions:

\begin{enumerate}
  \item \textbf{RQ1.} How much do per-peer client, architecture, hosting, and validator attributes change estimated node and network power compared with one wattage applied to every peer?
  \item \textbf{RQ2.} How do the model's predictions compare with wall-power traces from a reference machine?
\end{enumerate}

RQ1 needs a map from crawlable attributes to wattage, and a
check that a model trained on complete peers can fill the rest.
RQ2 needs wall-power traces that were not used to build those
wattages \citep{lei2021best,sai2024rigor}. The pipeline assigns
a wattage to each peer in a two-layer mainnet crawl
\citep{hackspacher2026ethereumcode}. Cloud attribution,
previously limited to Amazon Web Services (AWS)
\citep{pankovska2024electricity}, covers fourteen providers
through the Cloud Carbon Footprint (CCF) per-vCPU wattages
\citep{ccf_methodology}. ARM and x86 hosts receive separate
profiles. Validator activity is read per peer from attestation
subnets, sync-committee bits, and live gossip topics
\citep{eth2_p2p_spec,EIP778,eth2_altair_p2p}.
Section~\ref{sec:results} reports how well the forest recovers
the rule table, the snapshot total against a baseline that
applies the same wattage to every peer and against the CCAF
index, the split of watts by hosting and validator duty, and two
24-hour traces on a reference machine.

%% file: sections/03_related.tex
\section{Related Work}
\label{sec:related-work}

After the Merge, energy studies follow one of two paths. Some
scale a few measured wattages by a single network-wide headcount.
Others start from measured client and hardware pairs and then
guess how those pairs are distributed across the network. Hash
rate is gone, so the remaining draw is the wall power of the
machines that stay online
\citep{ccri2022merge,pankovska2024electricity}.
Table~\ref{tab:related-comparison} places the main models next to
the pipeline used here.

\begin{table}[ht]
\centering
\caption{Post-Merge Ethereum energy models.}
\label{tab:related-comparison}
\input{tables/tab_related_comparison.tex}
\end{table}

The Crypto Carbon Ratings Institute (CCRI) merge report
\citep{ccri2022merge} is the main measurement source. It reports
idle baselines and the extra watts each major consensus and
execution client adds on a few consumer machines, with both
clients synchronised to mainnet. Almost every later estimate,
including this one and the 2024 CCRI methods note
\citep{ccri2024whitepaper}, reuses those numbers. They are also
the only public set that covers all five then-common consensus
clients under one protocol. The cost of that reuse is inherited
bias: the same two or three chassis, and the same missing
clients, appear in every later model.

\citet{platt2021energy} published an earlier multi-chain
estimate. They assume one machine per validator, which pushes
totals up relative to models that allow several validators on
one host \citep{pankovska2024electricity}.
\citet{ibanez2023replication} put the CCRI figures into that
framework, confirm the post-Merge drop, and shrink the
uncertainty interval. Both still pick a global hardware mix
instead of reading the attributes of a named peer.

Cambridge's Blockchain Network Sustainability Index (CBNSI)
\citep{cbnsi_methodology} is the closest running estimate. It
pairs CCRI wattages with crawling: Armiarma from MigaLabs
characterises beacon nodes \citep{cortes2021discovering}, and
MonitorEth supplies counts for the hardware mix. That already
joins a crawl to a measurement table. ARM is still lumped with
x86, hosting is not read from a current IP-prefix list, and the
attestation and sync-committee bitvectors
(\texttt{attnets} and \texttt{syncnets}) are unused at the level
of individual peers. Those three fields are what move
per-node watts
\citep{cortes2023home,sutton2022validators}.
\citet{cortes2023home} show that a large share of the network
can run at home, including on small boards. One weighted wattage
cannot cover that low-power tail and a high-memory cloud
instance at the same time.

\citet{pankovska2024electricity} add the cloud-versus-home split
and a power-usage-effectiveness (PUE) factor that accounts for
data-centre overhead. Their model is the first to apply
instance-level Amazon Web Services (AWS) Elastic Compute Cloud
(EC2) figures to cloud nodes, with a 5~W solid-state-drive (SSD)
overhead and a cloud PUE of 1.2. We keep those two constants.
Their cloud path stops at AWS and still starts from monitoring
counts, so the virtualisation-aware formula is never applied
instance by instance. We map fourteen providers through Cloud
Carbon Footprint (CCF) per-vCPU wattages \citep{ccf_methodology},
keep ARM and x86 separate, and mark validators from ENR
bitvectors and live gossip topics instead of a fleet average
\citep{EIP778,eth2_p2p_spec}.

\citet{lei2021best} ask for measured hardware and an observed
machine mix, not nameplate watts or an assumed validator-to-host
ratio. \citet{sai2024rigor} treat a watt-meter check as required.
The node-level gaps above motivate RQ1; the call for a meter
check motivates RQ2. \citet{kiffer2025thirtysix} and
\citet{cortes2021discovering} already show who is online at peer
granularity, and libp2p DHT crawls attribute peers to cloud
providers by IP range \citep{trautwein2022ipfs}. The energy models
above still replace those peers with a mix. Outside blockchains,
\citet{hoffmann2026measureonce} follow the pattern we adopt:
benchmark under controlled conditions once, then evaluate compact
models on attributes observable at run time, without production
power telemetry. Research infrastructures face the same question,
from the MaRDI mathematical data portal \citep{giessing2024reducing}
to the data centres behind next-generation observatories
\citep{freour2025eso}; both propose efficiency measures but still
lack per-component measurements.

%% file: tables/tab_related_comparison.tex
\footnotesize
\setlength{\tabcolsep}{2.5pt}
\begin{tabularx}{\columnwidth}{@{}
  >{\raggedright\arraybackslash}p{1.28cm}
  >{\raggedright\arraybackslash}X
  >{\raggedright\arraybackslash}X
  >{\raggedright\arraybackslash}X
  >{\raggedright\arraybackslash}X
  @{}}
\toprule
 & CCRI / Platt & CCAF CBNSI & Pankovska et al. & This work \\
\midrule
Input
  & assumed global mix
  & crawl counts + mix
  & monitoring counts
  & per-peer crawl \\
Cloud scope
  & none / home split
  & none
  & AWS EC2 only
  & 14 providers \\
ARM vs.\ x86
  & lumped or fixed
  & lumped
  & separate paths
  & separate rules \\
Validator signal
  & fleet average
  & none at peer level
  & fleet average
  & ENR + gossipsub \\
Validation
  & CCRI lab
  & ---
  & ---
  & wall-power traces \\
\bottomrule
\end{tabularx}

%% file: sections/04_method.tex
\section{Method}
\label{sec:method}

The pipeline has four steps \citep{fayyad1996kdd}: collect a
two-layer crawl; extract client, architecture, OS, hosting, and
subnet features; assign wattage from published measurements when
those fields are complete; impute the rest with a forest trained
on the rule labels. Section~\ref{sec:empirical-protocol} covers
the wall-power setup for RQ2. Code and datasets are public
\citep{hackspacher2026ethereumcode}.

\subsection{Crawl and features}
\label{sec:crawl-collection}

We use two Nebula sessions \citep{nebula-github} on mainnet,
the crawler behind the IPFS DHT measurements of
\citet{trautwein2022ipfs}.
The CL crawl ran on 19~June~2026, 10:01--10:26~UTC
(232{,}831 visits). The EL crawl ran on 22~June~2026,
02:45--03:14~UTC (127{,}911 visits). Each visit is a
JSONL record with libp2p Identify fields \citep{libp2p_identify_spec},
the ENR \citep{EIP778}, and Nebula metadata. The visit dumps
are \texttt{out\_consensus.json} and
\texttt{out\_execution.json}.

IPv4 is taken from \texttt{connect\_maddr} \citep{libp2p_addressing};
records without it are dropped. Nebula only keeps peers that answer,
so DHT clients behind NAT that never accept inbound connections are
outside the set \citep{antswatch}; in libp2p networks, direct
connections to such peers depend on relay-assisted hole punching
\citep{trautwein2026dcutr}. Layers are inner-joined on
(address,~occurrence-index) so the $k$-th CL record at an address
matches the $k$-th EL record. We do not collapse
several records behind one public address into a single node:
one address can front several machines. Of 6{,}369 addresses
seen on both layers, 3{,}440 (54\%) have a different number of
consensus and execution records, so the $k$-th pairing is an
assumption rather than a guaranteed match.
Section~\ref{sec:network-total} reports the subset of addresses
that appear exactly once on each layer as a sensitivity check.
We keep the pairing rather than drop unmatched rows
\citep{christen2012datamatching}.
Rows with a null consensus-layer \texttt{fork\_digest} are
removed. That field marks mainnet membership on the consensus
layer and drops Optimism OP-Stack peers that reuse the same
networking stack \citep{eth2_p2p_spec,optimism_opstack}.
Table~\ref{tab:features} lists the normalised columns.

\begin{table}[ht]
\centering
\caption{Normalised crawl features used for wattage assignment
  and modelling.}
\label{tab:features}
\small
\begin{tabularx}{\columnwidth}{l >{\raggedright\arraybackslash}X}
\toprule
Feature & Source and handling \\
\midrule
\texttt{consensus\_client}
  & Consensus-layer \texttt{AgentVersion} matched against
    Lighthouse, Prysm, Teku, Nimbus, Lodestar, Grandine, Caplin. \\
\texttt{execution\_client}
  & Execution-layer match against Geth, Nethermind, Besu, Erigon,
    Reth, Nimbus-EL \citep{nimbus_eth1_docs}, Ethrex. \\
\texttt{hw\_arch}
  & \texttt{aarch64}/\texttt{arm64} versus
    \texttt{x86\_64}/\texttt{amd64}.
    Conflicting tokens are left unresolved. \\
\texttt{os\_token}
  & \texttt{linux}, \texttt{macos}, or \texttt{windows}.
    Conflicts are left unresolved. \\
\texttt{cloud\_provider}
  & IPv4 prefix match on AWS, GCP, Azure, Oracle, Hetzner, OVH,
    Contabo, Netcup, Latitude, DigitalOcean, Vultr, Linode, Leaseweb,
    Clouvider \citep{ripestat_api}. No match is unclassified, not
    proven bare metal. \\
\texttt{syncnets\_num}
  & Number of set bits in the Altair \texttt{syncnets} mask
    \citep{eth2_altair_p2p}. \\
\texttt{attnets\_num}
  & Number of set bits in the 64-bit \texttt{attnets} mask
    \citep{eth2_p2p_spec}. \\
\texttt{gossipsub\_att\_count}
  & Live \texttt{beacon\_attestation\_\{N\}} topics; used with the
    bitvectors to flag validators. \\
\bottomrule
\end{tabularx}
\end{table}

\subsection{Power assignment}
\label{sec:power-assignment}

A peer receives a rule-based wattage only if both clients, the
processor architecture, and the operating system can be
identified. Incomplete rows are passed to the Random Forest
(Section~\ref{sec:rf}). Caplin without Erigon is dropped because
that pairing is not a valid production configuration
\citep{caplin_erigon}. An empty hosting-provider field is
unclassified: the wattage rules treat it as a physical host
without a hypervisor. That choice overstates the home and
bare-metal share whenever a smaller cloud is missing from the
fourteen prefix lists.

\subsubsection{Bare-metal nodes}
\label{sec:bare-metal}

Unclassified prefixes are treated as physical hosts without a
hypervisor \citep{blockdaemon_bare_metal,cortes2023home}. We use
the CCRI idle baselines and the extra watts each client adds on
synchronised mainnet nodes
\citep{ccri2022merge,ccri2024whitepaper,cbnsi_methodology,pankovska2024electricity}.
When both clients share a host, the combined extra draw is
scaled by $f_{\text{adj}} = 0.91$ \citep{ccri2022merge}. Only
CCRI configurations~5 (Intel Core i5-10400, 64~GB, 25.04~W idle)
and~6 (AMD Ryzen Threadripper 3970X, 256~GB, 78.17~W idle)
remain plausible: current production guidance asks for 32~GB for
a non-validator and 64~GB for a validator
\citep{lighthouse_installation}.
Table~\ref{tab:ccri-tiers} records those two tiers.

\begin{table}[ht]
\centering
\caption{CCRI hardware configurations retained for unclassified
  x86 hosts \citep{ccri2022merge}.}
\label{tab:ccri-tiers}
\small
\begin{tabular}{clrr}
\toprule
Cfg. & CPU & RAM & Idle (W) \\
\midrule
5 & Intel Core i5-10400 & 64\,GB & 25.04 \\
6 & AMD Ryzen TR 3970X  & 256\,GB & 78.17 \\
\bottomrule
\end{tabular}
\end{table}

Six clients seen in the crawl have no CCRI measurement. Grandine
and Caplin take the Nimbus consensus-layer extra draw (the
lowest measured value, which is conservative for designs that
aim at low resource use). Nethermind, Nimbus-EL, and Ethrex take
the Geth figure; Reth takes Erigon
\citep{pathania2025software,grandine_github,caplin_announcement,cortes2023home,reth_github,nethermind_github,ethrex_github}.
The measured extra watts on configurations~5 and~6 follow
\citet{ccri2022merge}: Lighthouse 3.14/18.84~W, Prysm
2.87/24.33~W, Teku 3.32/27.46~W, Nimbus 2.08/17.11~W, Lodestar
3.89/33.55~W; Geth 9.70/47.70~W, Erigon 17.59/44.62~W, Besu
31.02/75.04~W.

For x86 Linux and Windows,
\begin{equation}
  P_{\text{node}}
  = \sum_{t \in \{5,6\}}
    w_t \cdot \bigl[(P_{\text{CL},t} + P_{\text{EL},t})
    \times f_{\text{adj}} + P_{\text{idle},t}\bigr],
  \label{eq:power-bare-metal}
\end{equation}
with $f_{\text{adj}} = 0.91$, $w_5 = 0.75$, and $w_6 = 0.25$.
CCRI originally weighted configurations~4, 5, and~6 as 0.25,
0.50, and 0.25. After dropping the under-specified tier~4, the
0.75/0.25 split matches the EthStaker survey in which 80\% of
home stakers run 64~GB or less \citep{ethstaker2025survey}.
Table~\ref{tab:bare-metal-rules} lists the other platforms. ARM
Windows is dropped: no RAPL or DCMI dataset exists for that pair
\citep{panigrahy2026energyblindspot,microsoft2025winhwreq}.

\begin{table}[ht]
\centering
\caption{Bare-metal wattage rules by architecture and operating
  system.}
\label{tab:bare-metal-rules}
\small
\begin{tabularx}{\columnwidth}{ll >{\raggedright\arraybackslash}X}
\toprule
Arch. & OS & Assignment \\
\midrule
x86 & Linux, Windows
  & Eq.~\eqref{eq:power-bare-metal}; CCRI idle
    (Table~\ref{tab:ccri-tiers}). \\
x86 & macOS
  & Same extra watts; idle $19.9$~W (2018 Intel Mac mini)
    \citep{apple_macmini_power}. \\
ARM & macOS
  & Same extra watts; idle $6.9$~W (M1/M2 Mac mini mean)
    \citep{apple_macmini_power}. \\
ARM & Linux
  & Fixed $10$~W \citep{web3pi2024validators,web3pi_about}. \\
ARM & Windows
  & Dropped. \\
\bottomrule
\end{tabularx}
\end{table}

The ARM Linux token names the instruction-set architecture, not
a specific system-on-chip. We apply the 10~W figure only because
it is the long-run public mainnet measurement on a
Raspberry~Pi~5; other ARM Linux boards are not treated as a
separate class.

\subsubsection{Cloud nodes}
\label{sec:cloud}

Cloud wattage depends on role
(Table~\ref{tab:node-hw-requirements}). The ENR
\texttt{attnets} field is a 64-bit mask of persistent
attestation-subnet subscriptions
\citep{EIP778,eth2_p2p_spec}. Before consensus-specs pull
request~3312, any non-zero mask marked a validator. Since
May~2023 every beacon node keeps two long-lived attestation
subnets derived from its node identity, so two set bits are the
non-validator baseline
\citep{eth2_attnets_pr3312,somoza2024blockprint}. Extra bits
still cost CPU. \citet{sutton2022validators} saw load rise with
each added attestation topic, because the node validates
unaggregated attestations on that subnet. The crawler also
reports live \texttt{beacon\_attestation\_\{N\}} gossip topics.
Only validator clients call the beacon
\texttt{committee\_subscriptions} API, so a topic count above
two is not explained by the backbone mandate
\citep{eth_beacon_api_validator_flow}. Any set bit in
\texttt{syncnets} is restricted to active validators
\citep{eth2_altair_p2p,eth2_altair_spec}.
A peer is flagged as a validator if
\texttt{attnets\_num}${}\,{>}\,2$,
\texttt{gossipsub\_att\_count}${}\,{>}\,2$, or
\texttt{syncnets\_num}${}\,{>}\,0$.
The two-subnet baseline applies only to attestation
subscriptions; any set sync-committee bit already marks an
active validator. The flag selects the hardware minima in
Table~\ref{tab:node-hw-requirements} for cloud nodes.
Unclassified x86 hosts still use the 0.75/0.25 mix of
configurations~5 and~6 regardless of duty; applying
configuration~6 only to validator-flagged unclassified nodes is
a sensitivity in Table~\ref{tab:sensitivity}.

\begin{table}[ht]
\centering
\caption{Minimum hardware assumed for a cloud node, by role
  \citep{lighthouse_installation,geth_hardware_requirements}.}
\label{tab:node-hw-requirements}
\small
\begin{tabular}{lrr}
\toprule
Resource & Non-validator & Validator \\
\midrule
RAM (GB) & 32 & 64 \\
vCPU     &  4 &  8 \\
SSD (TB) &  2 &  4 \\
\bottomrule
\end{tabular}
\end{table}

For AWS we follow \citet{pankovska2024electricity}:
\begin{equation}
  P_{\text{node}}^{\text{AWS}}
  = \bigl(P_{\text{pkg}} + P_{\text{RAM}} + \delta + P_{\text{SSD}}\bigr)
    \times \phi_{\text{AWS}},
  \label{eq:cloud-node-power}
\end{equation}
with $P_{\text{SSD}} = 5$~W and a data-centre overhead
$\phi_{\text{AWS}} = 1.2$.
$P_{\text{pkg}}+P_{\text{RAM}}+\delta$ is the Teads
100\%-utilisation figure \citep{teads2021ec2}, the closest
discrete level given typical node load above 50\%. We keep
m6i/r6i for x86 \citep{teads2021ec2,aws_rds_m6i_r6i}; r6g uses
Teads Graviton2 numbers and m7g a Graviton3 package term of
$100 \times n/64$~W
\citep{semianalysis_graviton3}. Candidates are
filtered by architecture and Table~\ref{tab:node-hw-requirements};
we pick the smallest qualifying instance
\citep{aws_rightsizing_2020}.

Other providers use the CCF component model
\citep{ccf_methodology}:
\begin{equation}
  P_{\text{node}}^{\text{non-AWS}}
  = \bigl(W_{\mathrm{avg}}\, n_{\text{vCPU}}
    + \beta_{\text{RAM}} n_{\text{RAM}}
    + \beta_{\text{SSD}} n_{\text{SSD}}\bigr)
    \times \phi_{\text{non-AWS}},
  \label{eq:non-aws-load-power}
\end{equation}
where $W_{\mathrm{avg}}=|M|^{-1}\sum_{m\in M}W_m$,
with $\beta_{\text{RAM}} = 0.6$~W\,GB$^{-1}$,
$\beta_{\text{SSD}} = 1.2$~W\,TB$^{-1}$, and
$\phi_{\text{non-AWS}} = 1.185$. $W_m$ is the per-vCPU maximum of
microarchitecture $m$; $M$ is the provider fleet mix (CCF and
SPECpower coefficients and prefix lists are in the public
repository
\citep{hackspacher2026ethereumcode,specpower}). Hardware counts
again follow Table~\ref{tab:node-hw-requirements}.

\subsection{Random Forest for incomplete peers}
\label{sec:rf}

Many peers advertise a hosting provider and subnet counts but
omit the client, architecture, or operating system. A
provider-only average would discard the validator bits, so we
fill incomplete rows with a Random Forest
\citep{chapelle2006ssl,sklearn_persistence,breiman2001randomforests}.
The model is a standard regressor and is not heavily tuned
\citep{shwartzziv2022tabular}. Inputs are the five categorical
columns plus the numeric subnet and gossip-topic counts. Empty
provider values become the unclassified token
\texttt{bare\_metal}. An \texttt{OrdinalEncoder} uses sentinels
$-1$ and $-2$; encoder and forest form one \texttt{Pipeline}
fitted only on the training split \citep{kaufman2012leakage}.
The labeled set is split 80/20; the forest has 500 trees.
We also record the out-of-bag $R^2$ and a five-fold
cross-validated $R^2$ on all labeled rows.

A hold-out $R^2$ against the rules, with client, architecture,
and operating system still visible, only shows that the trees
can rebuild a deterministic table. When the forest is applied we
hide \texttt{consensus\_client}, \texttt{execution\_client},
\texttt{hw\_arch}, and \texttt{os\_token} on the hold-out rows,
which matches the information available for incomplete peers.
Unresolved categoricals stay as their own encoded value
\citep{lenz2022imputation,caplin_erigon}. Caplin-without-Erigon
rows are invalid (six dropped).

\subsection{Empirical power measurement}
\label{sec:empirical-protocol}

We recorded 24-hour wall-power traces on a Windows~11 x86
machine for two synchronised mainnet full-node pairs and an idle
baseline (no validator keys). Pairing each load scenario with an
idle baseline on the same machine follows established software
energy measurement models
\citep{guldner2024measurement,hoffmann2025greenmetrics}. Table~\ref{tab:measurement-hw}
lists the chassis: a Dell Alienware Aurora~R16 gaming desktop.
Home stakers usually run headless boxes
\citep{ethstaker2025survey,cortes2023home}. The GPU idles near
33~W in a published review of this card
\citep{servethehome_rtx4090}; that figure was not measured on
this PSU. Clients do not require a discrete GPU
\citep{geth_hardware_requirements,ccri2022merge}. We report raw
and GPU-adjusted figures. A myStrom WiFi Switch sat inline with error margin under 1\% \citep{mystrom_wifi_switch}. Published windows
\citep{hackspacher2026ethereumcode}: Nimbus+Geth (11~July~2026),
Prysm+Nethermind (20--21~July~2026), idle (16--17~July~2026).
Measured idle is 71.1~W; after the 33~W GPU that is 38.1~W
against the pipeline's 38.3~W CCRI x86 idle
\citep{ccri2022merge,cbnsi_methodology}. Idle matches. The
traces test client load.

\begin{table}[H]
\centering
\caption{Gaming desktop used for the wall-power traces
  \citep{dell_aurora_r16,intel_i9_14900f}.}
\label{tab:measurement-hw}
\small
\begin{tabularx}{\columnwidth}{l >{\raggedright\arraybackslash}X}
\toprule
Component & Specification \\
\midrule
System & Dell Alienware Aurora~R16 \\
CPU & i9-14900F (8P+16E / 32 threads; 65~W base, 219~W turbo) \\
Memory & 64~GB DDR5-5600 \\
GPU & RTX~4090, 24~GB \\
PSU & 1000~W Platinum \\
OS / plan & Windows~11 Home; High Performance \\
\bottomrule
\end{tabularx}
\end{table}

Additionally, we explored the minimum hardware requirements required to interact with the Ethereum Blockchain. The appendix~\ref{app:helios} reports a six-hour Helios light-client
trace on a Raspberry~Pi Zero~2W. That measurement bounds the
draw of a remote-RPC light client. It is not meant to validate the
10~W ARM Linux full-node assignment \cite{web3pi2024validators}.

%% file: sections/05_results.tex
\section{Results}
\label{sec:results}

\subsection{Crawl coverage}
\label{sec:estimation-analysis}

Of 6{,}940 matched peers, 3{,}110 (44.8\%) receive a rule-based
label and 3{,}824 get a forest wattage (99.9\% coverage;
Table~\ref{tab:coverage}). Six Caplin/non-Erigon rows are
invalid.

\begin{table}[H]
\centering
\caption{Peer coverage for the 19/22~June~2026 crawl.}
\label{tab:coverage}
\input{tables/tab_coverage.tex}
\end{table}

\subsection{Network-level estimate}
\label{sec:network-total}

Summing the rule labels and the forest estimates on the 6{,}934
valid matched peers gives a reachable-set total of 415~kW, or
3.63~GWh if that draw were held for a year
(Table~\ref{tab:network-estimate}). Peers with a complete rule
label contribute 195~kW (mean 62.6~W); peers filled by the
forest contribute 220~kW (mean 57.5~W).

A uniform Lighthouse$+$Nethermind x86 assignment (the most
common labeled pair, with Nethermind on the Geth proxy) on the
same 6{,}934 peers is 431~kW (3.78~GWh/year). Per-peer mapping
lowers the total by 3.9\%. Client choice
moves the network sum only a little on this crawl; the weights
on the two hardware tiers move it more
(Section~\ref{sec:sensitivity}).

The snapshot is about 46\% of the CCAF/CBNSI figure of about
0.90~MW (7.87~GWh/year)
\citep{ccaf2026report,cbnsi_methodology}. At 59.8~W per peer,
CCAF's demand implies on the order of 15{,}000 nodes. Mean watts
are close: 59.8~W here versus about 60~W if CCAF's 0.90~MW is
spread over 15{,}000 nodes. Most of the gap is coverage. Peers
seen on only one layer, peers a crawl cannot dial, and churn
between the consensus-layer and execution-layer sessions, which
are three days apart, are all missing from our sum. If unseen
peers look like the observed mix, a straight scale-up would sit
near 0.85--0.90~MW. We do not apply that factor: there is no
passive count of nodes that never accept incoming connections
\citep{antswatch}.

\begin{table}[ht]
\centering
\caption{Estimated power of the reachable peers, compared with
  applying one wattage to every peer and with the CCAF/CBNSI
  index.}
\label{tab:network-estimate}
\input{tables/tab_network_estimate.tex}
\end{table}

Restricting the join to addresses that appear once on the scored
table leaves 4{,}898 peers and 297~kW. The 54\% of shared
addresses with a different number of consensus and execution
records therefore account for much of the snapshot uncertainty.

\subsection{Wattage shares}
\label{sec:compositional-effects}

On the scored set, unclassified prefixes are 52.6\% of peers and
54.3\% of watts; AWS is 12.2\% of peers and 15.6\% of watts;
Hetzner averages 42.9~W and accounts for 11.9\% of watts; and
validator-flagged peers are 25.7\% of the set and 31.4\% of
watts (Tables~\ref{tab:power-decomposition}
and~\ref{tab:power-decomposition-role}). AWS and
validator-flagged nodes use a larger share of watts than of
peers; Hetzner uses a smaller share.

\begin{table}[ht]
\centering
\caption{Share of peers and of estimated watts by hosting
  provider on the 6{,}934 scored peers.}
\label{tab:power-decomposition}
\input{tables/tab_power_decomposition.tex}
\end{table}

\begin{table}[ht]
\centering
\caption{Share of peers and of estimated watts by validator flag
  on the 6{,}934 scored peers.}
\label{tab:power-decomposition-role}
\input{tables/tab_power_decomposition_role.tex}
\end{table}

\begin{figure*}[!t]
\centering
\begin{subfigure}{0.47\textwidth}
\centering
\includegraphics[height=2.85cm,width=\linewidth,keepaspectratio]{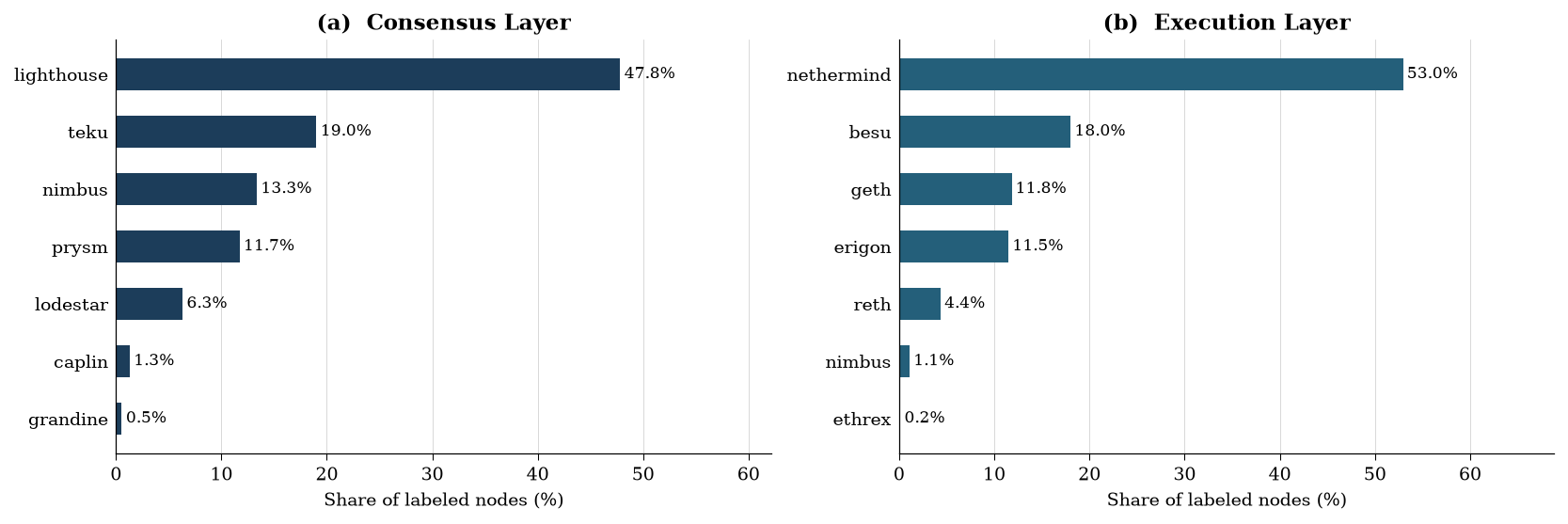}
\caption{Client software.}
\label{fig:client_distribution}
\end{subfigure}
\hfill
\begin{subfigure}{0.47\textwidth}
\centering
\includegraphics[height=2.85cm,width=\linewidth,keepaspectratio]{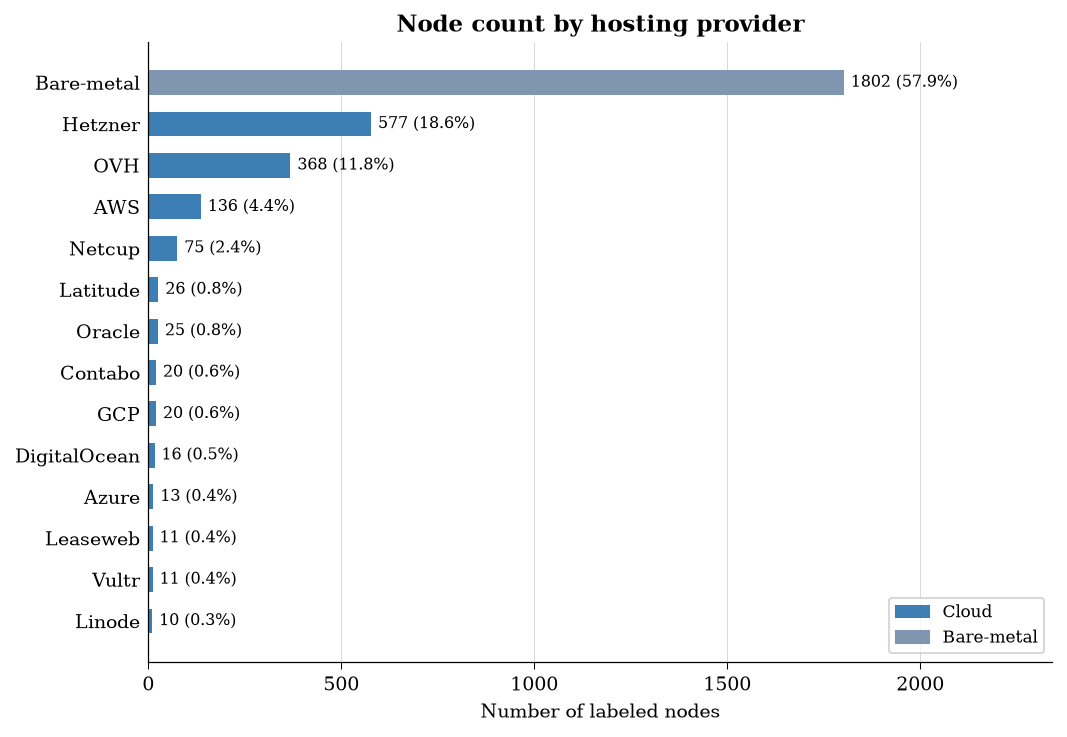}
\caption{Hosting provider.}
\label{fig:provider_distribution}
\end{subfigure}\\[0.25em]
\begin{subfigure}{0.47\textwidth}
\centering
\includegraphics[height=2.85cm,width=\linewidth,keepaspectratio]{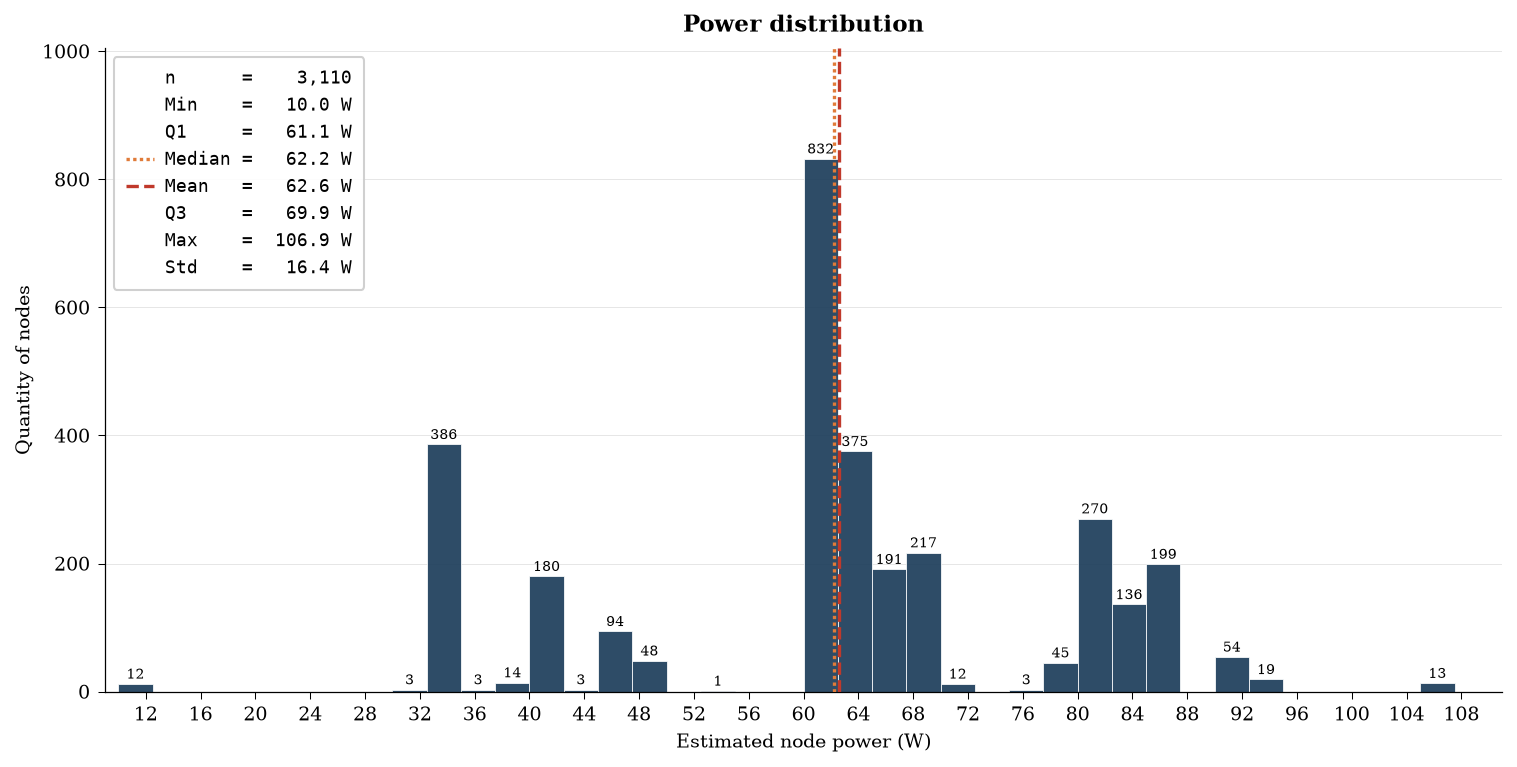}
\caption{Estimated per-node power.}
\label{fig:power_distribution}
\end{subfigure}
\hfill
\begin{subfigure}{0.47\textwidth}
\centering
\includegraphics[height=2.85cm,width=\linewidth,keepaspectratio]{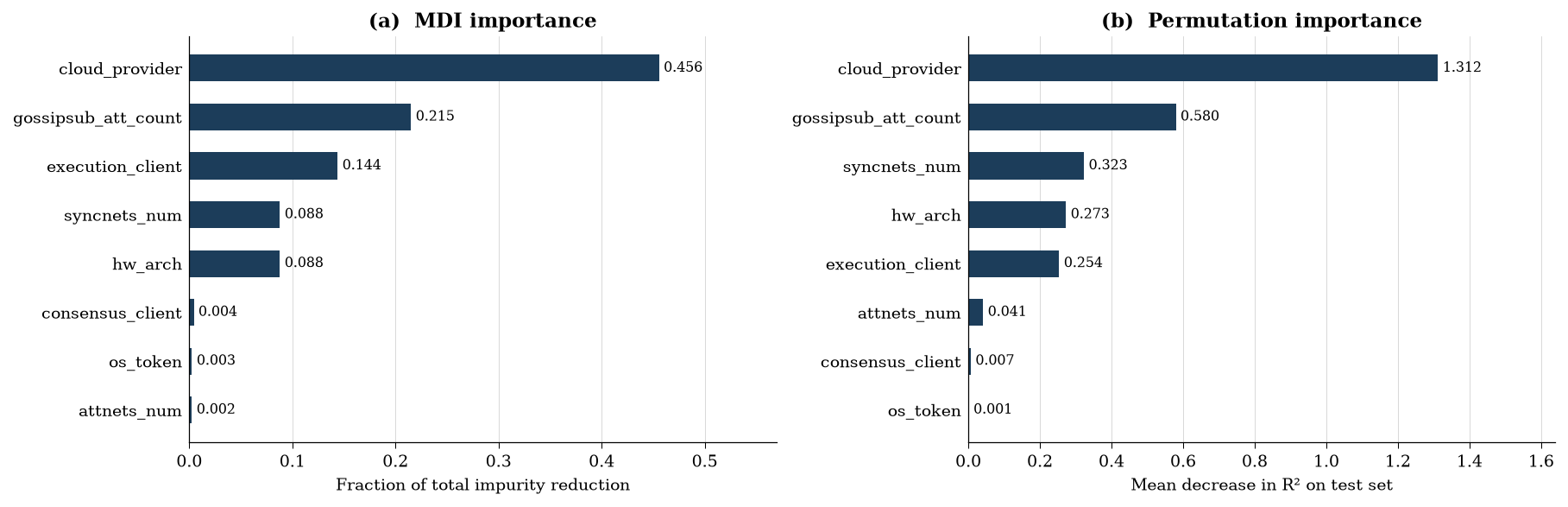}
\caption{Forest importances (MDI and permutation).}
\label{fig:feature_importances}
\end{subfigure}
\caption{Rule-based labeled mix ($n=3{,}110$): clients, hosting,
  power, and forest importances.}
\label{fig:client_provider}
\Description{Client and hosting bar charts, a power histogram peaked
  between 60 and 70 watts, and paired importance bars for cloud
  provider, gossipsub count, and execution client.}
\end{figure*}

Figure~\ref{fig:client_provider} shows the mix among peers that
received a rule-based label. On the consensus layer, Lighthouse
leads at 47.8\%
\citep{clientdiversityorg,ethereumorg_client_diversity}, then
Teku (19\%), Nimbus (13.3\%), Prysm (11.7\%), and Lodestar
(6.3\%); Caplin and Grandine stay below 2\%. On the execution
layer, Nethermind is 53\%, Besu 18\%, Geth 11.8\%, Erigon 11.5\%,
and Reth 4.4\%, with Nimbus-EL at 1.1\% and Ethrex at 0.2\%. Besu carries the highest CCRI execution extra
draw; almost one in five labeled nodes can therefore draw up to
three times the Geth/Nethermind proxy wattage. A majority of
labeled nodes (57.9\%) match no listed cloud prefix. Among
listed-cloud nodes, Hetzner is 18.6\% of the labeled set, OVH
11.8\%, and AWS 4.4\%; Hetzner and OVH together are 30.4\%.
Part of that European-cloud share is a vantage effect: the
crawler ran from Germany and can over-represent nearby hosts
\citep{kiffer2025thirtysix}.

Persistent attestation-subnet subscriptions spike at 2 (83.7\%
of labeled nodes); 6.1\% sit at 4 and 9.0\% subscribe to all 64
subnets (Table~\ref{tab:attnets_distribution}). The all-64 tail
validates unaggregated attestations on every subnet
\citep{sutton2022validators}. The wattage rules use those extra
bits only as a validator flag and to raise the cloud hardware
minima, so the extra CPU cycles are not in the wattage table.
Sync-committee subscriptions are mostly zero (79.0\%;
Table~\ref{tab:syncnets_distribution}). The remaining 13.3\%
with four set bits match nodes that hold, or are about to hold,
a sync-committee seat: the committee has 512 validators and
rotates about every 27 hours \citep{eth2_altair_spec}. Labeled power
concentrates between 60~W and 70~W (mean 62.6~W, median
62.2~W), with clusters at 32--48~W and 80--88~W and a tail to
106.9~W from AWS validator-tier instances
(Figure~\ref{fig:power_distribution}). The minimum is the 10~W
ARM Linux assignment.

Hosting provider dominates both importance measures (mean
decrease in impurity 0.456; permutation $\Delta R^2 = 1.312$),
then the live attestation-topic count (0.215 and 0.580).
Execution client is third by impurity decrease (0.144) but fifth
by permutation (0.254), because four of the seven execution
labels share the 9.70~W Geth proxy.
\texttt{syncnets\_num} and \texttt{hw\_arch} each contribute
0.088 of impurity decrease and rank third and fourth by
permutation (0.323 and 0.273). Consensus client, operating
system, and the attestation-bit count together contribute under
1\% of impurity decrease
(Figure~\ref{fig:feature_importances}).

\subsection{Sensitivity}
\label{sec:sensitivity}

Changing one assumption at a time on the labeled 195~kW
(Table~\ref{tab:sensitivity}) shows that the 0.75/0.25 hardware
mix dominates: flattening it to 0.50/0.50 raises the labeled
total by 24\%. Cloud PUE, AWS utilisation, and a
validator-aware mix for unclassified hosts move the labeled
total by a few percent. The ARM Linux 10~W point is negligible
on this snapshot.

\begin{table}[ht]
\centering
\caption{Labeled-set power under one-at-a-time assumption
  changes. Base is 195~kW.}
\label{tab:sensitivity}
\input{tables/tab_sensitivity.tex}
\end{table}

\subsection{Imputation}
\label{sec:model-evaluation}

On 622 held-out labeled nodes, with client, architecture, and
operating system still visible, the forest reaches a hold-out
root-mean-square error of 1.60~W, a mean absolute error (MAE)
of 0.28~W (0.44\% of the test-set mean), and
$R^2 = 0.9909$ (Table~\ref{tab:model-metrics}). Out-of-bag
$R^2$ is 0.9831; five-fold $R^2$ on all 3{,}110 labels is
$0.9890 \pm 0.0074$. Those scores measure agreement with the
rules, not with a watt-meter
\citep{shwartzziv2022tabular}. Hiding exactly the fields that
define a rule label, which is the setting used for incomplete
peers, drops hold-out $R^2$ to 0.77 and raises MAE to 4.3~W.
Hiding the hosting provider instead leaves $R^2 = 0.48$ and MAE
9.3~W, which matches the importance ranking: without a provider
the imputer is weak.
Figure~\ref{fig:pred_vs_actual} in the appendix is the
visible-field scatter.

\begin{table}[ht]
\centering
\caption{Random Forest evaluation against the rule wattages. The
  imputation row hides the fields that are required to create a
  rule label.}
\label{tab:model-metrics}
\input{tables/tab_model_metrics.tex}
\end{table}

\subsection{Empirical comparison}
\label{sec:empirical-comparison}

\begin{figure*}[!t]
\centering
\begin{subfigure}{0.47\textwidth}
\centering
\includegraphics[height=3.45cm,width=\linewidth,keepaspectratio]{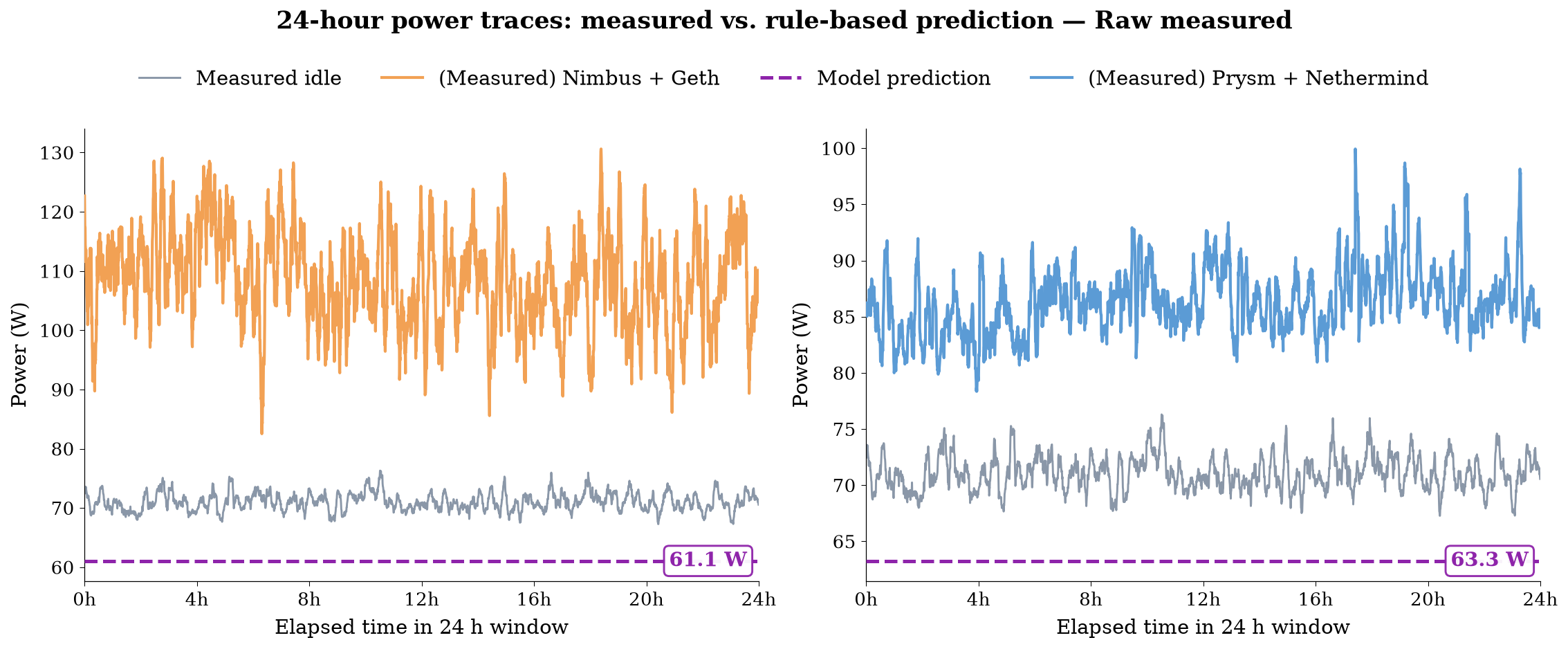}
\caption{Raw traces.}
\label{fig:empirical_comparison_raw}
\end{subfigure}
\hfill
\begin{subfigure}{0.47\textwidth}
\centering
\includegraphics[height=3.45cm,width=\linewidth,keepaspectratio]{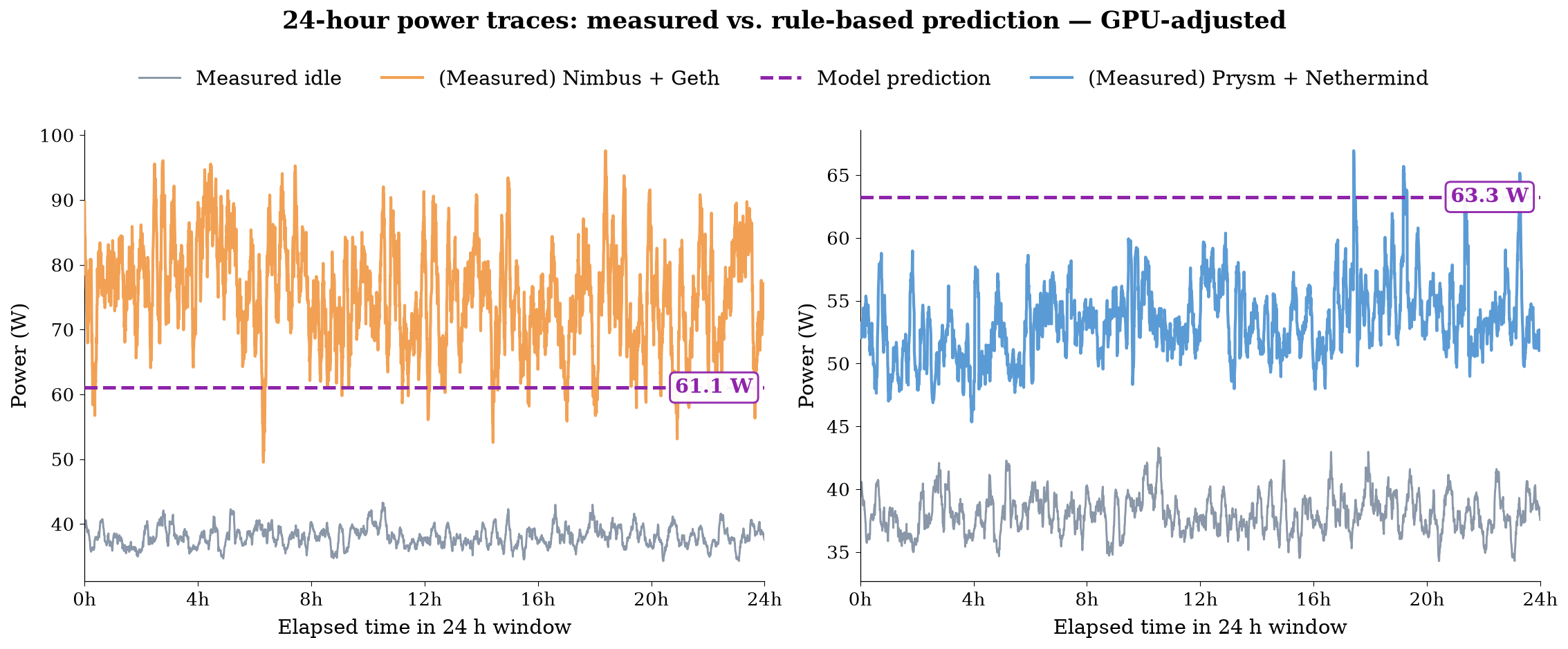}
\caption{Traces after subtracting idle GPU power.}
\label{fig:empirical_comparison_gpu_adjusted}
\end{subfigure}
\caption{24-hour wall-power traces compared with the model point
  predictions and the idle baseline.}
\label{fig:empirical_comparison}
\Description{Two 24-hour wall-power traces compared with horizontal
  model-prediction and idle lines, before and after subtracting GPU
  idle power.}
\end{figure*}

The gaming desktop is encoded as in
Table~\ref{tab:empirical-feature-row}: unclassified x86 Windows,
the two-subnet backbone, and the two client pairs. Traces use a
20-sample centred rolling average \citep{lei2021best}. Raw
traces sit above the model's point prediction all day. System
idle is 71.1~W; most of that is platform background. Subtracting
the 33~W RTX~4090 idle
\citep{servethehome_rtx4090,gsf2023measuring} leaves 38.1~W
idle and cuts the deviations from 43.7\% and 26.8\% to 19.0\%
and 18.5\%
(Figure~\ref{fig:empirical_comparison}, Table~\ref{tab:empirical-comparison}).
The leftovers have opposite signs ($-14.26$~W for Nimbus+Geth,
$+9.93$~W for Prysm+Nethermind). We do not add that residual to
415~kW as a constant offset.

\begin{table}[ht]
\centering
\caption{Feature rows submitted to the trained pipeline for the
  gaming desktop. Unclassified hosting is the pipeline token
  \texttt{bare\_metal}.}
\label{tab:empirical-feature-row}
\footnotesize
\setlength{\tabcolsep}{3pt}
\begin{tabularx}{\columnwidth}{@{}>{\raggedright\arraybackslash}X cc@{}}
\toprule
Feature & Nimbus+Geth & Prysm+Neth. \\
\midrule
\texttt{hw\_arch} / \texttt{os\_token}
  & x86 / windows & x86 / windows \\
\texttt{cloud\_provider} & unclassified & unclassified \\
\texttt{attnets\_num} & 2 & 2 \\
\texttt{gossipsub\_att\_count} & 2 & 2 \\
\texttt{syncnets\_num} & 0 & 0 \\
\bottomrule
\end{tabularx}
\end{table}

\begin{table}[ht]
\centering
\caption{Measured 24-hour mean power versus the model prediction
  on the bare-metal x86 Windows gaming desktop.}
\label{tab:empirical-comparison}
\input{tables/tab_empirical_comparison.tex}
\end{table}

%% file: tables/tab_coverage.tex
\small
\begin{tabularx}{\columnwidth}{@{}>{\raggedright\arraybackslash}X r@{}}
\toprule
Category & Count \\
\midrule
Total CL$+$EL inner-joined peers & 6{,}940 \\
Rule-based labeled               & 3{,}110 \\
Forest wattage                   & 3{,}824 \\
Invalid                          &      6 \\
\bottomrule
\end{tabularx}

%% file: tables/tab_network_estimate.tex
\small
\begin{tabularx}{\columnwidth}{@{}>{\raggedright\arraybackslash}X rr@{}}
\toprule
Estimate & Power & Annual energy \\
\midrule
This work (6{,}934 matched peers)
  & 415~kW & 3.63~GWh \\
Same peers, one Lighthouse and Nethermind x86 wattage
  & 431~kW & 3.78~GWh \\
CCAF/\mbox{CBNSI} (their census)
  & ${\sim}$0.90~MW & 7.87~GWh \\
\bottomrule
\end{tabularx}

%% file: tables/tab_power_decomposition.tex
\small
\setlength{\tabcolsep}{4pt}
\begin{tabularx}{\columnwidth}{@{}>{\raggedright\arraybackslash}X rrr@{}}
\toprule
Hosting & Nodes & Mean (W) & Share \\
\midrule
Unclassified (no listed prefix) & 3{,}645 & 61.7 & 54.3\% \\
AWS                             &   849 & 76.4 & 15.6\% \\
Hetzner                         & 1{,}149 & 42.9 & 11.9\% \\
OVH                             &   768 & 58.3 & 10.8\% \\
Other listed cloud              &   523 & 58.7 &  7.4\% \\
\bottomrule
\end{tabularx}

%% file: tables/tab_power_decomposition_role.tex
\small
\setlength{\tabcolsep}{4pt}
\begin{tabularx}{\columnwidth}{@{}>{\raggedright\arraybackslash}X rrr@{}}
\toprule
Role & Nodes & Mean (W) & Share \\
\midrule
Non-validator     & 5{,}152 & 55.2 & 68.6\% \\
Validator-flagged & 1{,}782 & 73.1 & 31.4\% \\
\bottomrule
\end{tabularx}

%% file: tables/tab_sensitivity.tex
\small
\setlength{\tabcolsep}{4pt}
\begin{tabularx}{\columnwidth}{@{}>{\raggedright\arraybackslash}X r@{}}
\toprule
Changed assumption & $\Delta P$ \\
\midrule
ARM Linux 8~W / 15~W
  & $-0.01$\% / $+0.03$\% \\
Equal 0.50/0.50 weights on configs~5 and~6
  & $+24.1$\% \\
Cloud PUE $1.1$ / $1.4$
  & $-2.8$\% / $+6.8$\% \\
AWS utilisation $50$\%
  & $-2.3$\% \\
Config~6 only for validator-flagged unclassified hosts
  & $+3.3$\% \\
\bottomrule
\end{tabularx}

%% file: tables/tab_model_metrics.tex
\small
\setlength{\tabcolsep}{3pt}
\begin{tabularx}{\columnwidth}{@{}>{\raggedright\arraybackslash}X rr@{}}
\toprule
Evaluation & $R^2$ & MAE (W) \\
\midrule
Hold-out vs.\ rules (fields visible)
  & 0.9909 & 0.28 \\
Five-fold CV vs.\ rules
  & $0.9890 \pm 0.0074$ & --- \\
Out-of-bag vs.\ rules
  & 0.9831 & --- \\
Hide client, arch, OS (imputation)
  & 0.77 & 4.34 \\
Hide hosting provider
  & 0.48 & 9.32 \\
\bottomrule
\end{tabularx}

%% file: tables/tab_empirical_comparison.tex
\footnotesize
\setlength{\tabcolsep}{3pt}
\begin{tabularx}{\columnwidth}{@{}>{\raggedright\arraybackslash}X cccc@{}}
\toprule
Client pair
  & GPU off$^{*}$
  & Meas.\ [W]
  & Pred.\ [W]
  & Err.\ [\%] \\
\midrule
Nimbus + Geth      & No  & 108.36 & 61.1 & 43.7 \\
Prysm + Nethermind & No  & 86.37  & 63.3 & 26.8 \\
Nimbus + Geth      & Yes & 75.36  & 61.1 & 19.0 \\
Prysm + Nethermind & Yes & 53.37  & 63.3 & 18.5 \\
\bottomrule
\end{tabularx}

\vspace{2pt}
{\footnotesize $^{*}$Subtract the RTX~4090 idle of 33~W
\citep{servethehome_rtx4090} from the 24-hour mean.}

%% file: sections/06_discussion.tex
\section{Discussion}
\label{sec:discussion}

Assigning a wattage from each peer's advertised attributes
changes the snapshot by $-3.9$\% compared with applying one
Lighthouse and Nethermind x86 wattage to every peer. AWS and
validator-flagged nodes use a larger share of watts than of
peers; Hetzner uses a smaller share. The 415~kW
(3.63~GWh/year) figure is the sum over peers that this crawl
can reach and match across both layers. It is a lower bound,
not a fleet count. It is about 46\%
of CCAF's ${\sim}$0.90~MW because this crawl sees fewer nodes;
mean watts stay near 60~W in both estimates. The hardware-tier
weights move the labeled total more than client choice does
($+24.1$\% if the 0.75/0.25 mix is flattened to 0.50/0.50).

That 3.63~GWh already exceeds the 0.5~GWh MiCAR threshold at
which a reporting entity must also fill in greenhouse-gas and
renewable-share fields \citep{micar2025rts}. Those fields apply
to a reporting entity, not to the network as a whole, and they
need a location and a grid mix that this paper does not
estimate. A host that claims 100\% renewable electricity still
counts in the energy total.

Besu, the highest-power CCRI execution client, is second at
18\% of the labeled execution-layer mix, so almost one in five
labeled nodes can draw up to three times the Geth/Nethermind
extra watts (Figure~\ref{fig:power_distribution}).
Validator-flagged peers average 73.1~W against 55.2~W for the
rest. The full consensus-client range on configuration~5 is
only 1.81~W \citep{ccri2022merge}, which is why the consensus
client contributes almost nothing to forest importance.
Hosting still moves the labeled histogram: Hetzner and OVH sit
in the lower half and pull the mode into 60--70~W, while AWS
accounts for the tail above 100~W through
Equation~\eqref{eq:cloud-node-power}. Validator signals
describe mostly non-staking full nodes (83.7\% at two
attestation subnets), a moderate cluster at three or four
subnets, and a 9\% all-subnet tail that pays the full
attestation-validation cost \citep{sutton2022validators};
13.3\% with four sync-committee bits match sync-committee
seats \citep{eth2_altair_spec}. The forest barely uses the
attestation-bit count because the rules do not add watts for
extra topics beyond the validator hardware step.

For RQ1, the visible-field forest scores agree because the
target is a deterministic rule table. Hold-out $R^2 = 0.9909$,
out-of-bag $R^2 = 0.9831$, and five-fold
$R^2 = 0.9890 \pm 0.0074$ all sit within 0.008 of one another.
Residuals concentrate at tier boundaries
(Figure~\ref{fig:pred_vs_actual})
\citep{athey2019machinelearning}. Those metrics are agreement
with the assignment rules, not with a meter. The scores that
matter for incomplete peers are the masked ones: hosting
recovers most of the label (MAE 4.3~W); without a provider the
imputer is weak.

For RQ2, raw idle on the test machine is 71.1~W. After removing
the RTX~4090, non-GPU idle is about 38.1~W against the
pipeline's 38.3~W x86 idle \citep{ccri2022merge}. More than
half of each raw deviation is that GPU offset, which no CCRI
box and no typical Ethereum node has
\citep{geth_hardware_requirements}. Remaining residuals have
opposite signs ($-14.26$~W for Nimbus+Geth, $+9.93$~W for
Prysm+Nethermind), so they come from the clients. Nethermind is
treated as lighter than Geth in other comparisons
\citep{pathania2025software}; the traces also ran on
Windows~11 while CCRI measured Linux
\citep{ccri2022merge,najmuddin2021linuxwindows}. We do not add
that residual to 415~kW as a constant. The validation chassis
is a 219~W-turbo gaming desktop with a 1000~W power supply;
home stakers report headless low-power boxes
\citep{ethstaker2025survey,cortes2023home}. The Helios 0.84~W
mean is light-client draw on a board that never stored the
chain. It is not an ARM full-node wattage.

\subsection{Limitations}
\label{sec:threats}

The rule wattages are published measurements and proxies, not a
meter on each peer. Forest scores against those rules are not
empirical accuracy. The consensus-layer and execution-layer
crawls are three days apart
\citep{grandjean2024participation,imtiaz2019churn,kiffer2025thirtysix},
and 54\% of shared addresses have a different number of records
on the two layers
\citep{heimbach2025deanonymizing,luo2025unveiling}. The 33~W GPU
correction is a published idle figure for the card, not a
measurement on this power supply. Six clients inherit proxy
wattages; on Windows, Nethermind drew less than its Geth proxy.
Process-level power attribution on Linux
\citep{hoffmann2025powerletrics} could replace these proxies with
per-client measurements on a shared host.

The two CCRI boxes set the unclassified x86 wattages
\citep{ccri2022merge}. The 0.75/0.25 mix moves the labeled
total by 24\% if flattened to 0.50/0.50. Hence, this distribution causes high sensitivity on the final energy consumption estimation. 
Unclassified prefixes include unlisted clouds. The crawl vantage is Germany. The
wall-power check covers one gaming desktop and two
non-validator pairs, with no cloud or ARM full node. The 33~W
GPU offset belongs to this chassis only.

Active crawls miss nodes that never accept inbound connections.
A passive keyspace listener such as Ants-Watch, run with the
same Nebula snapshot, can bound that hidden population from
incoming routing-table traffic \citep{antswatch}. We do not
scale 415~kW by an unseen factor. There is no count of how many
such peers exist. The 9\% all-attestation-subnet tail is extra
CPU on peers already in the table.

\balance

%% file: sections/07_conclusion.tex
\section{Conclusion}
\label{sec:conclusion}

RQ1 holds for the assignment table. Client identity,
architecture, operating system, hosting, and subnet activity
are enough to rebuild the wattages used here. A 500-tree forest
trained on 3{,}110 labels reaches hold-out $R^2 = 0.9909$,
five-fold $R^2 = 0.9890 \pm 0.0074$, and a normalised MAE of
0.44\%, and covers 99.9\% of matched peers. Those scores are
agreement with the rules, not with a watt-meter. When the
fields that define a rule label are hidden, hold-out MAE is
4.3~W.

On the same 6{,}934 matched peers, estimated draw is 415~kW, or
3.63~GWh if held for a year. That is 3.9\% below applying the
same Lighthouse and Nethermind x86 wattage to every peer, and
about 46\% of the CCAF index, mainly because this snapshot sees
fewer nodes. AWS and validator-flagged peers take a larger
share of watts than of headcount. Flattening the CCRI 0.75/0.25
hardware mix moves the labeled total by 24\%.

RQ2 is only partly closed. The traces come from one gaming
desktop. Raw idle is 71.1~W. After subtracting a 33~W
discrete-GPU idle, 24-hour means sit about 19\% from the
predictions and non-GPU idle matches the CCRI x86 baseline. The
leftover error is a proxied client and a Windows-versus-Linux
difference. 
Still missing are headless Linux traces, watts for the all-attestation-subnet tail, and a
same-day crawl with a passive listener for nodes that never
accept incoming connections \citep{antswatch}. 

From our measurements, we estimated that the Ethereum network, with measured 3.63~GWh/year is well above the 0.5~GWh/year threshold of the MiCAR regulation.
Therefore, service providers using Ethereum Tokens must provide full sustainability disclosures.
To comply, our results must be extended using grid mix data \citep{micar2023,micar2025rts}. 

Our whole estimation model is available at the Github repository: 
\url{https://github.com/yahncosta/ethereum-energy-code};
crawls, tables, and traces are at
\url{https://huggingface.co/datasets/yhackspacher/ethereum-energy-data}
\citep{hackspacher2026ethereumcode}.

%% file: sections/08_appendix.tex
\appendix
\section{Cloud assignment tables}
\label{app:cloud-tables}

Tables~\ref{tab:aws-variables} and~\ref{tab:ec2-instances}
define the symbols in Equation~\eqref{eq:cloud-node-power}.
Tables~\ref{tab:non-aws-variables}--\ref{tab:provider-microarchs}
supply the CCF coefficients used in
Equation~\eqref{eq:non-aws-load-power}.

\begin{table}[H]
\centering
\caption{Variables in the AWS node-power equation
  (Equation~\eqref{eq:cloud-node-power}).}
\label{tab:aws-variables}
\small
\begin{tabularx}{\columnwidth}{l >{\raggedright\arraybackslash}X r}
\toprule
Symbol & Definition & Value \\
\midrule
$P_{\text{pkg}}$    & CPU package power & Table~\ref{tab:ec2-instances} \\
$P_{\text{RAM}}$    & DRAM power & Table~\ref{tab:ec2-instances} \\
$\delta$            & Residual overhead & Table~\ref{tab:ec2-instances} \\
$P_{\text{SSD}}$    & Fixed SSD draw & 5~W \\
$\phi_{\text{AWS}}$ & Cloud PUE & 1.2 \\
\bottomrule
\end{tabularx}
\end{table}

\begin{table}[H]
\centering
\caption{EC2 instances retained for AWS node power estimation.}
\label{tab:ec2-instances}
\footnotesize
\setlength{\tabcolsep}{3pt}
\resizebox{\columnwidth}{!}{%
\input{tables/tab_ec2_instances.tex}}
\end{table}

\begin{table}[H]
\centering
\caption{Variables in the non-AWS cloud equation
  (Equation~\eqref{eq:non-aws-load-power}).}
\label{tab:non-aws-variables}
\small
\begin{tabularx}{\columnwidth}{l >{\raggedright\arraybackslash}X r}
\toprule
Symbol & Definition & Value \\
\midrule
$W_m$ & Per-vCPU max wattage
  & Tables~\ref{tab:hyperscaler-vcpu-w},~\ref{tab:ccf-microarch} \\
$M$ & Provider microarchitectures
  & Table~\ref{tab:provider-microarchs} \\
$n_{\text{vCPU}}$, $n_{\text{RAM}}$, $n_{\text{SSD}}$
  & Role-based minima & Table~\ref{tab:node-hw-requirements} \\
$\beta_{\text{RAM}}$ & RAM coefficient & 0.6~W\,GB$^{-1}$ \\
$\beta_{\text{SSD}}$ & SSD coefficient & 1.2~W\,TB$^{-1}$ \\
$\phi_{\text{non-AWS}}$ & Cloud PUE & 1.185 \\
\bottomrule
\end{tabularx}
\end{table}

\begin{table}[H]
\centering
\caption{Hyperscaler per-vCPU wattage.}
\label{tab:hyperscaler-vcpu-w}
\input{tables/tab_hyperscaler_vcpu_w.tex}
\end{table}

\begin{table}[H]
\centering
\caption{SPECpower-derived microarchitecture wattage
  \citep{specpower}.}
\label{tab:ccf-microarch}
\input{tables/tab_ccf_microarch.tex}
\end{table}

\begin{table}[H]
\centering
\caption{CPU microarchitectures per hosting provider (set $M$ in
  Equation~\eqref{eq:non-aws-load-power}).}
\label{tab:provider-microarchs}
\input{tables/tab_provider_microarchs.tex}
\end{table}

\section{Hold-out scatter}
\label{app:pred-vs-actual}

\begin{figure}[H]
\centering
\includegraphics[width=0.86\columnwidth]{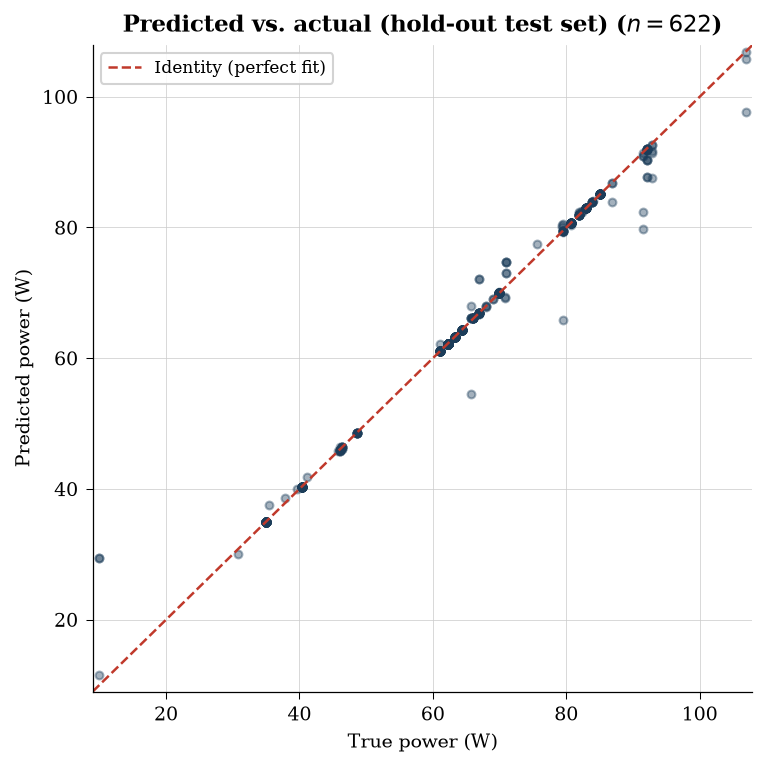}
\caption{Predicted versus rule-based power on 622 held-out nodes
  when the fields that define a rule label are visible.}
\label{fig:pred_vs_actual}
\Description{Scatter plot of predicted watts against rule-based labels
  near the identity line.}
\end{figure}

\section{Light-client bound}
\label{app:helios}

\begin{figure}[H]
\centering
\includegraphics[width=\columnwidth]{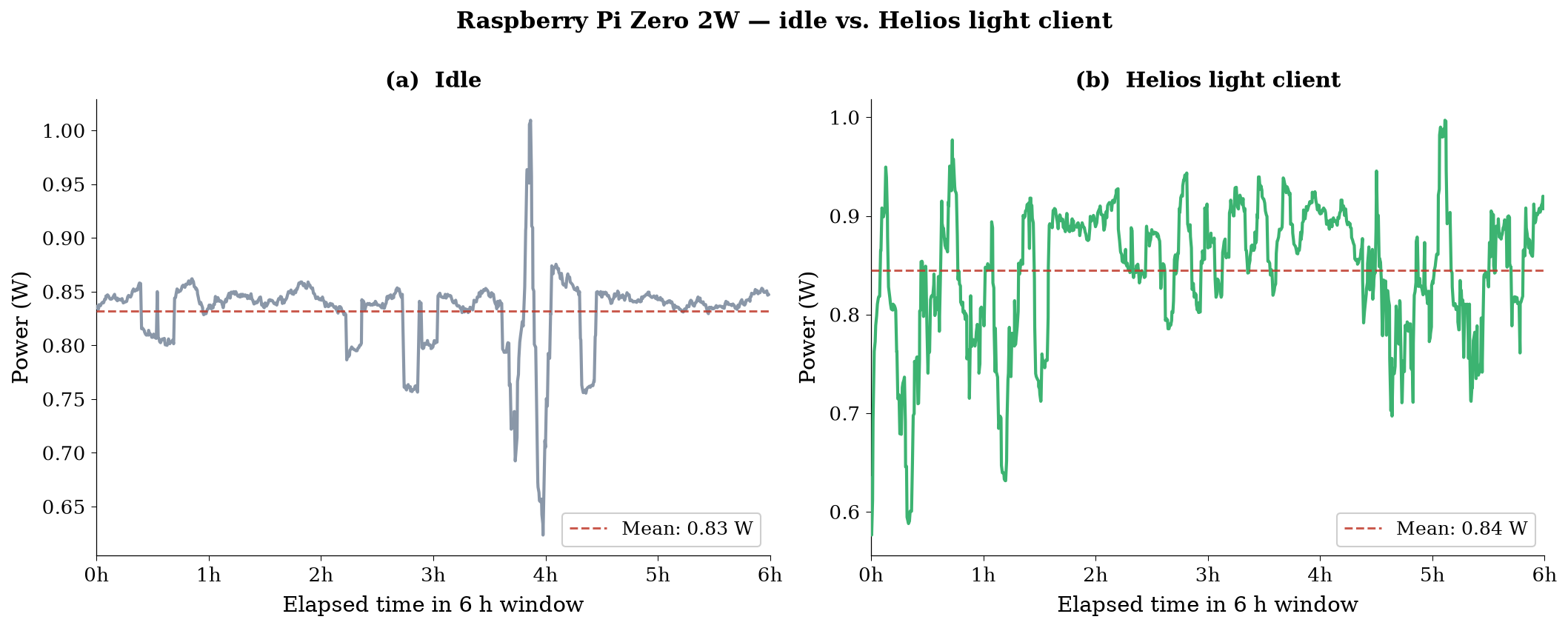}
\caption{Raspberry~Pi Zero~2W power over two six-hour windows:
  (a)~idle and (b)~Helios.}
\label{fig:raspberry_helios}
\Description{Two six-hour power traces on a Raspberry Pi Zero 2W,
  nearly overlapping at about 0.83 to 0.84 watts.}
\end{figure}

A Raspberry~Pi Zero~2W \citep{rpi_zero2w} was metered idle
(27~July) and under Helios (30~July), six hours each. Helios is
a light client: it checks headers against remote consensus
signatures and queries execution over RPC, and it does not store
local chain state \citep{helios_github,ethereum_light_clients}.
The board could not compile Helios; the binary was
cross-compiled for \texttt{aarch64}. Idle mean power is 0.83~W
and Helios is 0.84~W (Figure~\ref{fig:raspberry_helios}). The
0.01~W extra draw is the cost of a remote-RPC light client, not
of an ARM full node, and sits an order of magnitude below the
10~W ARM Linux full-node assignment
\citep{ethereum_light_clients,helios_github,web3pi2024validators}.

\section{Subnet subscription counts}
\label{app:subnet-tables}

\begin{table}[H]
\centering
\caption{Sync-committee subnet subscriptions.}
\label{tab:syncnets_distribution}
\input{tables/tab_syncnets_distribution.tex}
\end{table}

\begin{table}[H]
\centering
\caption{Attestation subnet subscriptions.}
\label{tab:attnets_distribution}
\input{tables/tab_attnets_distribution.tex}
\end{table}

\balance

%% file: tables/tab_ec2_instances.tex
\begin{tabular}{llrrrrrr}
\toprule
Instance & Arch & vCPU & RAM [GB] & $P_{\text{pkg}}$ [W] & $P_{\text{RAM}}$ [W] & $\delta$ [W] & $P_{\text{pct100}}$ [W] \\
\midrule
m6i.2xlarge & x86 & 8 & 32 & 38.20 & 19.20 & 7.50 & 64.90 \\
m6i.4xlarge & x86 & 16 & 64 & 76.39 & 38.40 & 15.00 & 129.79 \\
r6i.2xlarge & x86 & 8 & 64 & 38.20 & 38.40 & 7.50 & 84.10 \\
r6i.4xlarge & x86 & 16 & 128 & 76.39 & 76.80 & 15.00 & 168.19 \\
r6g.2xlarge & ARM & 8 & 64 & 19.10 & 38.40 & 3.75 & 61.20 \\
r6g.4xlarge & ARM & 16 & 128 & 38.20 & 76.80 & 7.50 & 122.50 \\
m7g.2xlarge & ARM & 8 & 32 & 12.50 & 19.20 & 3.80 & 35.50 \\
m7g.4xlarge & ARM & 16 & 64 & 25.00 & 38.40 & 7.50 & 70.90 \\
\bottomrule
\end{tabular}

%% file: tables/tab_hyperscaler_vcpu_w.tex
{\footnotesize
\begin{tabular}{lr}
\toprule
Provider & $W_m$ [W/vCPU] \\
\midrule
GCP   & 4.26 \\
Azure & 3.76 \\
\bottomrule
\end{tabular}
}

%% file: tables/tab_ccf_microarch.tex
{\footnotesize
\begin{tabular}{lr}
\toprule
Microarchitecture & $W_m$ [W/vCPU] \\
\midrule
EPYC 1st Gen     & 2.6042 \\
EPYC 2nd Gen     & 1.6930 \\
EPYC 3rd Gen     & 1.9573 \\
EPYC 4th Gen     & 2.2822 \\
EPYC 5th Gen     & 8.9614 \\
Cascade Lake     & 4.0632 \\
Ice Lake         & 3.7582 \\
Sapphire Rapids  & 4.1605 \\
Skylake          & 4.1042 \\
\bottomrule
\end{tabular}
}

%% file: tables/tab_provider_microarchs.tex
{\footnotesize
\setlength{\tabcolsep}{3pt}
\begin{tabularx}{\columnwidth}{l >{\raggedright\arraybackslash}X >{\raggedright\arraybackslash}X}
\toprule
Provider & CPU Microarchitectures & Sources \\
\midrule
Hetzner & EPYC 2nd, 3rd, 4th Gen & \citet{hetzner_ccx_press,hetzner_ax162} \\
OVH & EPYC 3rd, 4th Gen; Cascade Lake; Sapphire Rapids & \citet{ovh_cpu,ovh_adv_gen3} \\
Contabo & EPYC 2nd, 4th, 5th Gen & \citet{contabo_genoa,contabo_turin_vps} \\
Netcup & EPYC 2nd, 4th, 5th Gen & \citet{netcup_cpu} \\
DigitalOcean & Skylake; Cascade Lake; Ice Lake; Sapphire Rapids; EPYC 2nd, 3rd Gen & \citet{do_premium,do_plan} \\
Vultr & Cascade Lake; EPYC 2nd, 3rd Gen & \citet{vultr_cpu,vultr_epyc3} \\
Linode & EPYC 1st, 2nd, 3rd Gen & \citet{linode_cpu} \\
Leaseweb & Cascade Lake; Sapphire Rapids; EPYC 2nd, 3rd Gen & \citet{leaseweb_cpu,leaseweb_epyc} \\
Clouvider & Cascade Lake; EPYC 3rd Gen & \citet{clouvider_cpu} \\
Latitude & EPYC 3rd, 4th, 5th Gen & \citet{latitude_cpu,latitude_pricing} \\
Oracle & EPYC 3rd, 4th, 5th Gen & \citet{oracle_cpu} \\
\bottomrule
\end{tabularx}
}

%% file: tables/tab_syncnets_distribution.tex
\begin{tabular}{rrr}
\toprule
\texttt{syncnets\_num} & Nodes & Share [\%] \\
\midrule
0 & 2{,}457 & 79.0\% \\
1 & 139 & 4.5\% \\
2 & 67 & 2.2\% \\
3 & 34 & 1.1\% \\
4 & 413 & 13.3\% \\
\midrule
Total & 3{,}110 & 100.0\% \\
\bottomrule
\end{tabular}

%% file: tables/tab_attnets_distribution.tex
\begin{tabular}{rrr}
\toprule
\texttt{attnets\_num} & Nodes & Share [\%] \\
\midrule
0 & 32 & 1.0\% \\
2 & 2{,}602 & 83.7\% \\
3 & 5 & 0.2\% \\
4 & 190 & 6.1\% \\
64 & 281 & 9.0\% \\
\midrule
Total & 3{,}110 & 100.0\% \\
\bottomrule
\end{tabular}